\documentstyle[12pt,axodraw]{article}

\begin{document} 
\pagestyle{empty}

\begin{center} 
{\Large {\bf A supersymmetric 3-4-1 model}}
\end{center}

\begin{center}
M. C. Rodriguez \\
{\it Funda\c c\~ao Universidade Federal do Rio Grande-FURG \\
Departamento de F\'\i sica \\
Av. It\'alia, km 8, Campus Carreiros \\
96201-900, Rio Grande, RS \\
Brazil}
\end{center}

\begin{abstract}  
We build the complete supersymmetric version of a 3-4-1 gauge model
using the superfield formalism. 
We point out that a discrete symmetry, similar to the R-symmetry in the 
minimal supersymmetric standard model, is possible to be defined in this 
model. Hence we have both R-conserving and R-violating possibilities. We 
also discuss some phenomenological results coming from this model.  
\end{abstract}


\section{Introduction}
\label{sec:intro}

The full symmetry of the so called Standard Model (SM) is the gauge group
$SU(3)_{c}\otimes SU(2)_{L}\otimes U(1)_{Y}$. Nevertheless, the SM is 
not considered as the ultimate theory since neither the fundamental 
parameters, masses and couplings, nor the symmetry pattern are 
predicted. Even though many aspects of the SM are experimentally 
supported to a very accuracy, the embedding of the model into a 
more general framework is to be expected.

Some of these possibilities is that, at energies of a few TeVs, the gauge
symmetry may be $SU(3)_{c}\otimes SU(3)_{L}\otimes U(1)_{N}$ 
(3-3-1 for shortness) \cite{pp,pf,mpp}. Recently, the supersymmetric version of these 
model have alreday benn constructed in \cite{331susy,331susy2}. These 3-3-1 models can be embedded in a 
model with 3-4-1, 
its mean $SU(3)_{c}\otimes SU(4)_{L}\otimes U(1)_{N}$ gauge symmetry 
\cite{su4b}. 

In $SU(4)_L\otimes U(1)_N$, the most general expression for the electric charge generator is a linear combination of the four diagonal
generators of the gauge group
\begin{eqnarray}
\label{ch}
Q&=&\frac{1}{2} \left( a \lambda_{3}+\frac{b}{\sqrt{3}} \lambda_{8}+ \frac{c}{\sqrt{6}} \lambda_{15} \right)+ NI_{4 \times 4} \nonumber \\
&=&\mbox{diag} \left[ \frac{1}{2} \left( a+ \frac{b}{3}+ \frac{c}{6}  \right)+N, 
\frac{1}{2} \left( -a+ \frac{b}{3}+ \frac{c}{6}  \right)+N, \frac{1}{2} \left( \frac{-2b}{3}+ \frac{c}{6}  \right)+N, 
\frac{-c}{4}+N \right], \nonumber \\
\label{chargeral}
\end{eqnarray} 
where $\lambda_{i}$, being the Gell-Mann matrices
for $SU(4)_L$, see \cite{greiner,su4}, normalized as  Tr$(\lambda_i\lambda_j)=2\delta_{ij}$,
$I_{4 \times 4}= \mbox{diag}(1,1,1,1)$ is the diagonal $4\times 4$ unit matrix, and $a$, $b$ 
and $c$ are free parameters to be fixed next. Therefore, there is an infinite number of models can, 
in principle, be constructed.

 A model with the
$SU(4)\otimes U(1)$ symmetry in the lepton sector, quarks were not considered on this work, was suggested some
years ago in Ref.~\cite{su4a}, in wchich the magnetic moment of neutrinos arises as the result of 
charged scalars that belong to an $SU(4)$ sextet, and the mass of neutrino arises at two-loop level 
as the result of electroweak radiative correction. 

The 3-4-1 model in Ref.~\cite{su4b} contain exotic electric charges 
only in the quark sector, while leptons have ordinary electric charges and gauge bosons have 
integer electric charges. The best feature of this model is that it provides us with an alternative to 
the problem of the number $N_f$ of fermion families. These sort of models are
anomaly free only if there are equal number of quadriplet and anti-quadriplet (considering the color 
degrees of freedom), and
furthermore requiring
the sum of all fermion charges to vanish. Two of the three quark
generations transform identically and one generation, it does not
matter which one, transforms in a different representation of
$SU(4)_L  \otimes U(1)_N$. This means that in these models
as in the
$SU(3)_c\otimes SU(3)_L\otimes U(1)_N$ ones~\cite{pp},  in order
to cancel anomalies, the number of families $(N_f)$ must be divisible
by the number of color degrees of freedom ($n$). This fact, together with asymptotic freedom in QCD, 
the model predicts that the number of generations must be three and only three.

On the other hand, at low energies these models
are indistinguishable from the SM. 
There is a very nice review about this kind of model see \cite{col,ind}. This make 3-4-1 model
interesting by their own. In this article we construct the supersymmetric version of the model 
in Ref~\cite{su4b}. 

The outline of the paper is as follows. In Sec.~\ref{sec:model} we present the
representation content of the supersymmetric 3-4-1 model. 
We build the lagrangian in Sec.~\ref{sec:lagrangian}. While in Sec.~\ref{sec:double}, 
we discuss the double charged charginos inthis model, while in the last section we 
present our conclusion.

\section{The model}
\label{sec:model}

In this section (Sec.~\ref{subsec:nonsusy}) we review the non-supersymmetric
3-4-1 model of Refs.~\cite{su4b} and add the 
superpartners (Sec.~\ref{subsec:susyp}) of the usual particles of the non supersimmetric model. 
The superfields, useful to construct the supersimmetric lagrangian of the model, associated with the particles 
of this model are introduced in section (Sec.~\ref{subsec:superfields}). 
 
\subsection{The representation content}
\label{subsec:nonsusy}

In the model of Ref.~\cite{su4b}, the free parameters for the eletric charge generators 
are
\begin{equation}
a=1, \quad b=-1, \quad c=-4, 
\end{equation}
and Eq.(\ref{chargeral}) can be rewritten as
\begin{eqnarray}
Q&=&\frac{1}{2} \left( \lambda_{3}-\frac{1}{\sqrt{3}}\lambda_{8}-\frac{4}{\sqrt{6}}\lambda_{15} 
\right) +NI_{4 \times 4}, \nonumber \\
&=&\mbox{diag}(N,N-1,N,N+1).
\label{q}
\end{eqnarray}

However, let us first consider the particle content
of the model without supersymmetry. We have the leptons
transforming in the lowest representation of $SU(4)_{L}$ the quartet\footnote{In the same way as proposed by 
Voloshin \cite{su4a} in order to understand the existence of neutrinos with large magnetic moment and small mass.} in the following way
\begin{equation}
L_{aL}= \left( 
\begin{array}{c}
\nu_{a}\\
l_{a}\\
\nu^{c}_{a}\\
l^{c}_{a}
\end{array}
\right)_L \sim ( {\bf 1},{\bf 4},0), \,\ a=1,2,3.
\label{trip}
\end{equation}
In parenthesis it appears the transformations properties under the respective
factors $(SU(3)_C,SU(4)_L,U(1)_N)$.

In the quark sector, one quark family is also put in the quartet 
representation 
\begin{eqnarray}
 Q_{1L} = \left(
\begin{array}{c}
u_{1}\\
d_{1}\\
u^{\prime}\\
J
\end{array}\right)_L \sim \left( {\bf 3},{\bf 4}, \frac{2}{3} \right) , 
\label{q1l}
\end{eqnarray}
and the respective singlets are given by
\begin{eqnarray}
u^{c}_{1L} &\sim& \left({\bf3}^*,{\bf1},-\frac{2}{3}\right),\quad
d^{c}_{1L} \sim \left({\bf3}^*,{\bf1},\frac{1}{3}\right),\nonumber \\ 
u^{\prime c}_{L} &\sim& \left({\bf3}^*,{\bf1},-\frac{2}{3}\right),\quad 
J^{c}_{L} \sim \left({\bf3}^*,{\bf1},-\frac{5}{3}\right),
\label{q1r}
\end{eqnarray}
writing all the fields as left-handed; $u^{\prime}$ and $J$ are new quarks with charge $+2/3$
and $+5/3$ respectively.

The others two quark generations, as we have explained in the introduction, we put in the anti-quartet 
representation
\begin{equation}
\begin{array}{cc}  
Q_{2 L} = \left(
\begin{array}{c}
d_{2}\\
u_{2}\\
d^{\prime}_{1}\\
j_{1}
\end{array}\right)_L \sim \left({\bf3},{\bf4}^{*},- \frac{1}{3}\right) ,\quad  
 
Q_{3L} = 
      \left( \begin{array}{c} 
d_{3}\\
u_{3}\\
d^{\prime}_{2}\\
j_{2}
\end{array} \right)_{L} 
\sim \left({\bf3},{\bf4}^{*},- \frac{1}{3}\right) , 
\end{array}
\label{q23l}
\end{equation}
and also with the respective singlets,
\begin{eqnarray}
u^{c}_{\alpha L} &\sim& \left({\bf3}^*,{\bf1},-\frac{2}{3}\right),\quad
d^{c}_{\alpha L} \sim \left({\bf3}^*,{\bf1},\frac{1}{3}\right), \nonumber \\
d^{\prime c}_{\beta L} &\sim& \left({\bf3}^*,{\bf1},\frac{1}{3}\right), \quad
j^{c}_{\beta L}  \sim \left({\bf3}^*,{\bf1},\frac{4}{3}
\right) \,\ , 
\label{q23r}
\end{eqnarray}
$j_{\beta}$ and $d^{\prime}_{\beta}$, $\beta =1,2$ are new quarks with charge $-4/3$ and
$-1/3$ respectively, while $\alpha =2,3$ is the familly index for the quarks. We remind that in Eqs.~(\ref{trip},\ref{q1l},\ref{q1r},\ref{q23l},\ref{q23r})
all fields are still symmetry eigenstates.

On the other hand, the scalars, in quartet, which are necessary to generate the 
quark masses are
\begin{eqnarray} 
\eta&=&\left(
\begin{array}{c}
\eta_{1}^{0} \\ \eta^{-}_{1} \\ \eta^{0}_{2} \\ \eta^{+}_{2}
\end{array}
\right) \sim \left( {\bf 1},{\bf 4},0 \right) \nonumber \\
\rho&=&\left(
\begin{array}{c}
\rho_{1}^{+} \\ \rho^{0} \\ \rho^{+}_{2} \\ \rho^{++}
\end{array}
\right) \sim \left( {\bf 1},{\bf 4},1 \right) \nonumber \\
\chi&=&\left(
\begin{array}{c}
\chi_{1}^{-} \\ \chi^{--} \\ \chi^{-}_{2} \\ \chi^{0}
\end{array}
\right) \sim \left( {\bf 1},{\bf 4}, -1 \right).
\label{3t} 
\end{eqnarray}
In order to avoid mixing among primed and unprimed
quarks, we have to introduce 
an extra scalar transforming like $\eta$ but with different 
vacuum expectation value (VEV)
\begin{equation}
\phi=\left(
\begin{array}{c}
\phi_{1}^{0} \\ \phi^{-}_{1} \\ \phi^{0}_{2} \\ \phi^{+}_{2}
\end{array}
\right) \sim \left( {\bf 1},{\bf 4},0 \right) \,\ .
\end{equation} 
In order to obtain massive charged leptons it is necessary to introduce 
the following symmetric anti-decuplet
\begin{equation}
H=\left(
\begin{array}{cccc}
H^{0}_{1}\, &\, H^{+}_{1}\, &\, H_{2}^{0}\, &\, H_{2}^{-} \\
H^{+}_{1}\, &\, H_{1}^{++}\, &\, H_{3}^{+}\, &\, H_{3}^{0} \\
H^{0}_{2}\, &\, H^{+}_{3}\, &\, H^{0}_{4}\, &\, H^{-}_{4} \\
H^{-}_{2}\, &\, H^{0}_{3}\, &\, H^{-}_{4}\, &\, H^{--}_{2}
\end{array}
\right) \sim \left( {\bf 1},{\bf 10}^{*},0 \right).
\label{sextet}
\end{equation}
then the charged leptons get a mass but
neutrinos remain massless, at least at tree level.

\subsection{Supersymmetric partners}
\label{subsec:susyp}

Now, we introduce the minimal set of particles in order to implement the
supersymmetry~\cite{mssm}. We have the sleptons corresponding to the leptons in
Eq.~(\ref{trip}); squarks related to the quarks in Eqs.(\ref{q1r})-(\ref{q23r});
and the Higgsinos related to the scalars given in Eqs.~(\ref{3t}) and
(\ref{sextet}). Then, we have to introduce the following additional particles 
\begin{eqnarray}
\tilde{Q}_{1L} &=& \left(
\begin{array}{c}
\tilde{u}_{1}\\
\tilde{d}_{1}\\
\tilde{u}^{\prime}\\
\tilde{J}
\end{array}\right)_L \sim \left({\bf3},{\bf4},\frac{2}{3}\right), \qquad 
\tilde{Q}_{\alpha L} = \left(
\begin{array}{c}
\tilde{d}_{\alpha} \\
\tilde{u}_{\alpha}\\
\tilde{d}^{\prime}_{\beta}\\
\tilde{j}_{\beta}
\end{array}\right)_L \sim \left( {\bf 3},{\bf 4}^{*},- \frac{1}{3} \right), \nonumber \\
\tilde{L}_{aL}&=&\left(\begin{array}{c}
\tilde{\nu}_{a}\\
\tilde{l}_{a}\\
\tilde{\nu}^{c}_{a}\\
\tilde{l}^{c}_{a}
\end{array}\right)_L \sim ( {\bf 1},{\bf 4},0) , \nonumber \\
u^{c}_{iL} &\sim& \left({\bf3}^*,{\bf1},-\frac{2}{3}\right),\quad
d^{c}_{iL} \sim \left({\bf3}^*,{\bf1},\frac{1}{3}\right),\nonumber \\ 
u^{\prime c}_{L} &\sim& \left({\bf3}^*,{\bf1},-\frac{2}{3}\right),\quad 
J^{c}_{L} \sim \left({\bf3}^*,{\bf1},-\frac{5}{3}\right), \nonumber \\
d^{\prime c}_{ \beta L} &\sim& \left({\bf3}^*,{\bf1},\frac{1}{3}\right), \quad
j^{c}_{ \beta L} \sim \left({\bf3}^*,{\bf1},\frac{4}{3}
\right) \,\ .
\label{lq}
\end{eqnarray}
Where $\alpha=2,3$ and $\beta=1,2$. The higgsinos of these model are given by
\begin{eqnarray}
\tilde{\eta}&=&\left(
\begin{array}{c}
\tilde{\eta}_{1}^{0} \\ \tilde{\eta}^{-}_{1} \\ \tilde{\eta}^{0}_{2} \\ 
\tilde{\eta}^{+}_{2}
\end{array}
\right) \sim \left( {\bf 1},{\bf 4},0 \right), \quad
\tilde{\phi}=\left(
\begin{array}{c}
\tilde{\phi}_{1}^{0} \\ \tilde{\phi}^{-}_{1} \\ \tilde{\phi}^{0}_{2} \\ 
\tilde{\phi}^{+}_{2}
\end{array}
\right) \sim \left( {\bf 1},{\bf 4},0 \right), \nonumber \\
\tilde{\rho}&=&\left(
\begin{array}{c}
\tilde{\rho}_{1}^{+} \\ \tilde{\rho}^{0} \\ \tilde{\rho}^{+}_{2} \\ 
\tilde{\rho}^{++}
\end{array}
\right) \sim \left( {\bf 1},{\bf 4},1 \right), \quad
\tilde{\chi}=\left(
\begin{array}{c}
\tilde{\chi}_{1}^{-} \\ \tilde{\chi}^{--} \\ \tilde{\chi}^{-}_{2} \\ 
\tilde{\chi}^{0}
\end{array}
\right) \sim \left( {\bf 1},{\bf 4},-1 \right), \nonumber \\
\end{eqnarray}
\begin{eqnarray}
\tilde{H}=\left(
\begin{array}{cccc}
\tilde{H}^{0}_{1}\, &\, \tilde{H}^{+}_{1}\, &\, \tilde{H}_{2}^{0}\, &\, 
\tilde{H}_{2}^{-} \\
\tilde{H}^{+}_{1}\, &\, \tilde{H}_{1}^{++}\, &\, \tilde{H}_{3}^{+}\, &\, 
\tilde{H}_{3}^{0} \\
\tilde{H}^{0}_{2}\, &\, \tilde{H}^{+}_{3}\, &\, \tilde{H}^{0}_{4}\, &\, 
\tilde{H}^{-}_{4} \\
\tilde{H}^{-}_{2}\, &\, \tilde{H}^{0}_{3}\, &\, \tilde{H}^{-}_{4}\, &\, 
\tilde{H}^{--}_{2}
\end{array}
\right) \left( {\bf 1},{\bf 10}^{*},0 \right).
\label{}
\end{eqnarray}

Besides, in order to to cancel chiral anomalies generated by the
superpartners of the scalars, we have to add the following higgsinos in the
respective anti-quartet representation, 
\begin{eqnarray} 
\eta^{\prime}&=&\left(
\begin{array}{c}
\eta_{1}^{\prime 0} \\ \eta^{\prime +}_{1} \\ \eta^{\prime 0}_{2} \\ 
\eta^{\prime -}_{2}
\end{array}
\right) \sim \left( {\bf 1},{\bf 4}^{*},0 \right), \quad
\phi^{\prime}=\left(
\begin{array}{c}
\phi_{1}^{\prime 0} \\ \phi^{\prime +}_{1} \\ \phi^{\prime 0}_{2} \\ 
\phi^{\prime -}_{2}
\end{array}
\right) \sim \left( {\bf 1},{\bf 4}^{*},0 \right) \nonumber \\
\rho^{\prime}&=&\left(
\begin{array}{c}
\rho_{1}^{\prime -} \\ \rho^{\prime 0} \\ \rho^{\prime -}_{2} \\ 
\rho^{\prime --}
\end{array}
\right) \sim \left( {\bf 1},{\bf 4}^{*},-1 \right), \quad
\chi^{\prime}=\left(
\begin{array}{c}
\chi_{1}^{\prime +} \\ \chi^{\prime ++} \\ \chi^{\prime +}_{2} \\ 
\chi^{\prime 0}
\end{array}
\right) \sim \left( {\bf 1},{\bf 4}^{*},1 \right),
\label{shtc}  
\end{eqnarray}
and the decuplet
\begin{equation}
H^{\prime}=\left(
\begin{array}{cccc}
H^{\prime 0}_{1}\, &\, H^{\prime -}_{1}\, &\, H_{2}^{\prime 0}\, &\, 
H_{2}^{\prime +} \\
H^{\prime -}_{1}\, &\, H_{1}^{\prime --}\, &\, H_{3}^{\prime -}\, &\, 
H_{3}^{\prime 0} \\
H^{\prime 0}_{2}\, &\, H^{\prime -}_{3}\, &\, H^{\prime 0}_{4}\, &\, 
H^{\prime +}_{4} \\
H^{\prime +}_{2}\, &\, H^{\prime 0}_{3}\, &\, H^{\prime +}_{4}\, &\, 
H^{\prime ++}_{2}
\end{array}
\right) \sim \left( {\bf 1},{\bf 10},0 \right).
\label{shsc} 
\end{equation}
Their superpartners, higgsinos, are
\begin{eqnarray}
\tilde{\eta}^{\prime}&=&\left(
\begin{array}{c}
\tilde{\eta}_{1}^{\prime 0} \\ \tilde{\eta}^{\prime +}_{1} \\ 
\tilde{\eta}^{\prime 0}_{2} \\ \tilde{\eta}^{\prime -}_{2}
\end{array}
\right) \sim \left( {\bf 1},{\bf 4}^{*},0 \right), \quad
\tilde{\phi}^{\prime}=\left(
\begin{array}{c}
\tilde{\phi}_{1}^{\prime 0} \\ \tilde{\phi}^{\prime +}_{1} \\ 
\tilde{\phi}^{\prime 0}_{2} \\ \tilde{\phi}^{\prime -}_{2}
\end{array}
\right) \sim \left( {\bf 1},{\bf 4}^{*},0 \right), \nonumber \\
\tilde{\rho}^{\prime}&=&\left(
\begin{array}{c}
\tilde{\rho}_{1}^{\prime -} \\ \tilde{\rho}^{\prime 0} \\ 
\tilde{\rho}^{\prime -}_{2} \\ \tilde{\rho}^{\prime --}
\end{array}
\right) \sim \left( {\bf 1},{\bf 4}^{*},-1 \right), \quad
\tilde{\chi}^{\prime}=\left(
\begin{array}{c}
\tilde{\chi}_{1}^{\prime +} \\ \tilde{\chi}^{\prime ++} \\ 
\tilde{\chi}^{\prime +}_{2} \\ \tilde{\chi}^{\prime 0}
\end{array}
\right) \sim \left( {\bf 1},{\bf 4}^{*},1 \right),
\label{}
\end{eqnarray}
\begin{equation}
\tilde{H}^{\prime}=\left(
\begin{array}{cccc}
\tilde{H}^{\prime 0}_{1}\, &\, \tilde{H}^{\prime -}_{1}\, &\, 
\tilde{H}_{2}^{\prime 0}\, &\, \tilde{H}_{2}^{\prime +} \\
\tilde{H}^{\prime -}_{1}\, &\, \tilde{H}_{1}^{\prime --}\, &\, 
\tilde{H}_{3}^{\prime -}\, &\, \tilde{H}_{3}^{\prime 0} \\
\tilde{H}^{\prime 0}_{2}\, &\, \tilde{H}^{\prime -}_{3}\, &\, 
\tilde{H}^{\prime 0}_{4}\, &\, \tilde{H}^{\prime +}_{4} \\
\tilde{H}^{\prime +}_{2}\, &\, \tilde{H}^{\prime 0}_{3}\, &\, 
\tilde{H}^{\prime +}_{4}\, &\, \tilde{H}^{\prime ++}_{2}
\end{array}
\right) \sim \left( {\bf 1},{\bf 10},0 \right).
\label{}
\end{equation}

The vev of our scalars are given by
\begin{eqnarray}
\langle\eta\rangle&=& \left( \frac{v}{\sqrt{2}},0,0,0 \right), \,\ 
\langle\rho\rangle= \left( 0, \frac{u}{\sqrt{2}},0,0 \right), \,\
\langle\phi\rangle= \left( 0,0,\frac{z}{\sqrt{2}},0 \right), \,\ 
\langle\chi\rangle= \left( 0,0,0, \frac{w}{\sqrt{2}} \right), \nonumber \\ 
\langle H_3^0\rangle&=& \frac{x}{\sqrt{2}}, \,\ \langle H^0_{1} \rangle= \langle H^0_{2} \rangle = \langle H^0_{4}\rangle=0, \nonumber \\
\langle\eta^{\prime}\rangle&=& \left( \frac{v^{\prime}}{2},0,0,0 \right), \,\ 
\langle\rho^{\prime}\rangle= \left( 0, \frac{u^{\prime}}{\sqrt{2}},0,0 \right), \,\
\langle\phi^{\prime}\rangle= \left( 0,0,\frac{z^{\prime}}{\sqrt{2}},0 \right), \,\ 
\langle\chi\rangle= \left( 0,0,0, \frac{w^{\prime}}{\sqrt{2}} \right), 
\nonumber \\
\langle H_3^{\prime 0}\rangle&=& \frac{x^{\prime}}{\sqrt{2}}, \,\ 
\langle H^{\prime 0}_{1} \rangle= \langle H^{\prime 0}_{2} \rangle = \langle H^{\prime 0}_{4}\rangle=0. \,\  \nonumber \\
\label{vev}
\end{eqnarray}

Concerning the gauge
bosons and their superpartners, if we denote the gluons by $g^b$ the respective
superparticles, the gluinos, are denoted by $\lambda^b_{C}$, with 
$b=1, \ldots,8$; and in the electroweak sector we have
$V^{a}$, with $a=1, \ldots, 15$; the gauge boson of $SU(4)_{L}$, and their gauginos partners  
$\lambda^b_{A}$; finally we have the gauge boson of 
$U(1)_{N}$, denoted by $V^\prime$, and its supersymmetric partner $\lambda_{B}$.

\subsection{Superfields}
\label{subsec:superfields}

The superfields formalism is useful in writing the Lagrangian which is 
manifestly invariant under the supersymmetric transformations~\cite{wb} with 
fermions and scalars put in chiral superfields while the gauge bosons in 
vector superfields. As usual the superfield of a field $\phi$ will be denoted
by $\hat{\phi}$~\cite{mssm}.
The chiral superfield of a multiplet $\phi$ is denoted by 
\begin{eqnarray}
\hat{\phi}\equiv\hat{\phi}(x,\theta,\bar{\theta})&=& \tilde{\phi}(x) 
+ i \; \theta \sigma^{m} \bar{ \theta} \; \partial_{m} \tilde{\phi}(x) 
+\frac{1}{4} \; \theta \theta \; \bar{ \theta}\bar{ \theta} \; \Box 
\tilde{\phi}(x) \nonumber \\ 
& & \mbox{} +  \sqrt{2} \; \theta \phi(x) 
+ \frac{i}{ \sqrt{2}} \; \theta \theta \; \bar{ \theta} \bar{ \sigma}^{m}
\partial_{m}\phi(x)                   
\nonumber \\ && \mbox{}+  \theta \theta \; F_{\phi}(x), 
\label{phi}
\end{eqnarray}
while the vector superfield is given by
\begin{eqnarray}
\hat{V}(x,\theta,\bar\theta)&=&-\theta\sigma^m\bar\theta V_m(x)
+i\theta\theta\bar\theta
\overline{\tilde{V}}(x)-i\bar\theta\bar\theta\theta
\tilde{V}(x)\nonumber \\ &+&\frac{1}{2}\theta\theta\bar\theta\bar\theta D(x).
\label{vector}
\end{eqnarray}
The fields $F$ and $D$ are auxiliary fields which are needed to
close the supersymmetric algebra and eventually will be eliminated using 
their motion equations. 

Summaryzing, we have in the 3-4-1 supersymmetric model the following 
superfields:
$\hat{L}_{1,2,3}$, $\hat{Q}_{1,2,3}$, $\hat{\eta}$, $\hat{\rho}$, 
$\hat{\chi}$, $\hat{\phi}$, $\hat{H}$; $\hat{\eta}^\prime$, $\hat{\rho}^\prime$, 
$\hat{\chi}^\prime$, $\hat{\phi}^{\prime}$, $\hat{H}^\prime$;
$\hat{u}^c_{1,2,3}$, $\hat{d}^c_{1,2,3}$, $u^{\prime}$, $d^{\prime}_{1,2}$, 
$\hat{J}$ and $\hat{j}_{1,2}$, i.e., 28 chiral
superfields, and 24 vector superfields: $\hat{V}^a$, $\hat{V}^\alpha$ and
$\hat{V}^\prime$. 

\section{The Lagrangian}
\label{sec:lagrangian}

With the superfields introduced in the last section we can built a 
supersymmetric invariant lagrangian. It has the following form
\begin{equation}
   {\cal L}_{341} = {\cal L}_{SUSY} + {\cal L}_{soft}.
\label{l1}
\end{equation}
Here ${\cal L}_{SUSY}$ is the supersymmetric piece, while ${\cal L}_{soft}$ 
explicitly breaks SUSY.
Below we will write each of these lagrangians in terms of the 
respective superfields. 

\subsection{The Supersymmetric Term.}
\label{subsec:st}

The supersymmetric term can be divided as follows
\begin{equation} 
   {\cal L}_{SUSY} =   {\cal L}_{Lepton}
                  + {\cal L}_{Quarks} 
                  + {\cal L}_{Gauge} 
                  + {\cal L}_{Scalar}, 
\label{l2}
\end{equation}

where each term is given by

 \begin{equation} 
  {\cal L}_{\mbox{L\'epton}} 
= \int d^{4}\theta\;\left[\,\hat{ \bar{L}}e^{2g\hat{V}} 
 \hat{L} \,\right], 
\label{l3}
\end{equation}

\begin{eqnarray}
{\cal L}_{\mbox{Quarks}} 
&=& \int d^{4}\theta\;\left \{ 
\hat{ \bar{Q}}_{1}e^{ \left[ 2g_{s}\hat{V}_{C}+2g\hat{V}+g^{\prime}
\left( \frac{2}{3}\right) \hat{V}^{\prime} \right]} \hat{Q}_{1}+
\,\hat{\bar{Q}}_{\alpha}e^{ \left[ 2g_{s}\hat{V}_{C}+2g\hat{\bar{V}}+g^{\prime}
\left( - \frac{1}{3} \right) \hat{V}^{\prime} \right]} \hat{Q}_{\alpha} \right. \nonumber \\
&+&\left. 
\hat{ \bar{u}}^{c}_{i}e^{ \left[ 2g_{s} \hat{ \bar{V}}_{C}+g^{\prime}
\left( - \frac{2}{3}\right) \hat{V}^{\prime} \right]} \hat{u}^{c}_{i}+ 
\hat{ \bar{d}}^{c}_{i}e^{ \left[ 2g_{s} \hat{ \bar{V}}_{C}+g^{\prime} 
\left( \frac{1}{3}\right)\hat{V}^{\prime} \right]} \hat{d}^{c}_{i} 
+ \hat{ \bar{J}}^{c}e^{ \left[ 2g_{s} \hat{ \bar{V}}_{C}+g^{\prime}
\left( - \frac{5}{3}\right)\hat{V}^{\prime} \right]} \hat{J}^{c} \right. \nonumber \\ 
&+&\left. 
\hat{ \bar{j}}^{c}_{ \beta}e^{ \left[ 2g_{s} \hat{ \bar{V}}_{C}+g^{\prime}
\left( \frac{4}{3}\right)\hat{V}^{\prime} \right]} \hat{j}^{c}_{ \beta} 
+ \hat{ \bar{u}}^{\prime c}e^{[2g_{s} \hat{ \bar{V}}_{C}+g^{\prime} \left( - \frac{2}{3}\right) \hat{V}']} \hat{u}^{\prime c} 
+\hat{ \bar{d}}^{\prime c}_{\beta}e^{[2g_{s} \hat{ \bar{V}}_{C}+g^{\prime} \left( \frac{1}{3}\right) \hat{V}']} \hat{d}^{\prime c}_{\beta}
\right\}, \nonumber \\
\label{l4}
\end{eqnarray}
where $\alpha=2,3$ and $\beta=1,2$, while the third term is
\begin{eqnarray} 
  {\cal L}_{\mbox{Gauge}} 
&=&  \frac{1}{4} \int  d^{2}\theta\;Tr[W_{C}W_{C}]+ \frac{1}{4} \int  d^{2}\theta\;Tr[W_{L}W_{L}]+
\frac{1}{4} \int  d^{2}\theta W^{ \prime}W^{ \prime} \nonumber \\ &+&  
\frac{1}{4} \int  d^{2}\bar{\theta}\;Tr[\bar{W}_{C}\bar{W}_{C}]+ \frac{1}{4} \int  d^{2}\bar{\theta}\;Tr[\bar{W}_{L}\bar{W}_{L}]+
\frac{1}{4} \int  d^{2}\bar{\theta} \bar{W}^{ \prime}\bar{W}^{ \prime} \nonumber \\
\label{l5}
\end{eqnarray}
where $\hat{V}_{c}=T^{a}\hat{V}^{a}_{c}$, 
$\hat{V}=T^{i}\hat{V}^{i}$ and $T^a=\lambda^{a}/2$ are the generators of 
$SU(3)$ i.e., $a=1,\cdots,8$, and $T^i=\lambda^{i}/2$ are the generators of 
$SU(4)$ i.e., $i=1,\cdots,15$, and $g_{s}$, $g$ and $g^{\prime}$ are the gauge 
coupling of $SU(3)_{C}$, $SU(4)_{L}$ and $U(1)_N$. $W^{a}_{c}$, $W^{i}$ and $W^{ \prime}$ 
are the strength fields, and they are given by 
\begin{eqnarray}
W^{a}_{\alpha c}&=&- \frac{1}{8g_{s}} \bar{D} \bar{D} e^{-2g_{s} \hat{V}_{c}} 
D_{\alpha} e^{-2g_{s} \hat{V}_{c}} \nonumber \\
W^{a}_{\alpha}&=&- \frac{1}{8g} \bar{D} \bar{D} e^{-2g \hat{V}} 
D_{\alpha} e^{-2g \hat{V}} \nonumber \\
W^{\prime}_{\alpha}&=&- 
\frac{1}{4} \bar{D} \bar{D} D_{\alpha} \hat{V}^{\prime} \,\ .
\label{l6}
\end{eqnarray}

Finally 
\begin{eqnarray}
  {\cal L}_{\mbox{Escalar}} 
&=& \int d^{4}\theta\;\left[\,\hat{ \bar{ \eta}}e^{2g\hat{V}} \hat{ \eta} + 
\hat{ \bar{ \rho}}e^{ \left( 2g\hat{V}+g^{\prime}\hat{V}^{\prime} \right)} \hat{ \rho} +
\hat{ \bar{ \chi}}e^{ \left( 2g\hat{V}-g^{\prime}\hat{V}^{\prime} \right)} \hat{ \chi} +
\hat{ \bar{ \phi}}e^{2g\hat{V}} \hat{ \phi} + 
\hat{ \bar{H}}e^{2g\hat{V}} \hat{H} \right. \nonumber \\
&+& \left.\,\hat{ \bar{ \eta}}^{\prime}e^{2g\hat{ \bar{V}}} \hat{ \eta}^{\prime} + 
\hat{ \bar{ \rho}}^{\prime}e^{ \left( 2g\hat{ \bar{V}}-g^{\prime}\hat{V}^{\prime} \right)} \hat{ \rho}^{\prime} +
\hat{ \bar{ \chi}}^{\prime}e^{ \left( 2g\hat{ \bar{V}}+g^{\prime}\hat{V}^{\prime} \right)} \hat{ \chi}^{\prime} + 
\hat{ \bar{ \phi}}^{\prime}e^{2g\hat{V}} \hat{ \phi}^{\prime} +
\hat{ \bar{H}}^{\prime}e^{2g\hat{ \bar{V}}} \hat{H}^{\prime} \right] 
\nonumber \\
&+& \int d^{2}\theta W+ \int d^{2}\bar{ \theta}\overline{W}
\label{l7}
\end{eqnarray}
where $W$ is the superpotential, which we discuss in the next subsection. 

\subsection{Superpotential.}
\label{subsec:spotential}

The superpotential of our model is given by
\begin{equation}
W=\frac{W_{2}}{2}+ \frac{W_{3}}{3}, 
\label{sp1}
\end{equation}
with $W_{2}$ having only two chiral superfields and the terms permitted by  
our symmetry are
\begin{equation}
W_{2}=\sum_{a=1}^{3}\mu_{0a}\hat{L}_{aL} \hat{ \eta}^{\prime}+
\sum_{a=1}^{3}\mu_{1a}\hat{L}_{aL} \hat{ \phi}^{\prime}+ 
\mu_{ \eta} \hat{ \eta} \hat{ \eta}^{\prime}+
\mu_{ \phi} \hat{ \phi} \hat{ \phi}^{\prime}+
\mu_{2} \hat{ \eta} \hat{ \phi}^{\prime}+
\mu_{3} \hat{ \phi} \hat{ \eta}^{\prime}+
\mu_{ \rho} \hat{ \rho} \hat{ \rho}^{\prime}+ 
\mu_{ \chi} \hat{ \chi} \hat{ \chi}^{\prime}+
\mu_{H} \hat{H} \hat{H}^{\prime},
\label{sp2}
\end{equation}
and in the case of three chiral superfields the terms are
\begin{eqnarray}
W_{3}&=& \sum_{a=1}^{3} \sum_{b=1}^{3} \sum_{c=1}^{3}
\lambda_{1abc} \epsilon \hat{L}_{aL} \hat{L}_{bL} \hat{L}_{cL}+
\sum_{a=1}^{3} \sum_{b=1}^{3}
\lambda_{2ab} \epsilon \hat{L}_{aL} \hat{L}_{bL} \hat{ \eta}+
\sum_{a=1}^{3} \sum_{b=1}^{3}
\lambda_{3ab} \epsilon \hat{L}_{aL} \hat{L}_{bL} \hat{ \phi} \nonumber \\
&+&
\sum_{a=1}^{3} \sum_{b=1}^{3} 
\lambda_{4ab} \hat{L}_{aL} \hat{L}_{bL} \hat{H}  
+ \sum_{a=1}^{3} \lambda_{5a} \epsilon \hat{L}_{aL} \hat{\chi} \hat{\rho}+
f_{1} \epsilon \hat{ \rho} \hat{ \chi} \hat{ \eta}+
f_{2} \epsilon \hat{ \rho} \hat{ \chi} \hat{ \phi}+
f_{3} \hat{ \eta} \hat{ \eta} \hat{H}+
f_{4} \hat{ \eta} \hat{ \phi} \hat{H} \nonumber \\ &+&
f_{5} \hat{ \phi} \hat{ \phi} \hat{H}+
f_{6} \hat{ \chi} \hat{ \rho} \hat{H}+ 
f^{\prime}_{1}\epsilon \hat{\rho}^{\prime}\hat{\chi}^{\prime}\hat{\eta}^{\prime}+
f^{\prime}_{2}\epsilon \hat{\rho}^{\prime}\hat{\chi}^{\prime}\hat{\phi}^{\prime}+
f^{\prime}_{3}\hat{\eta}^{\prime}\hat{\eta}^{\prime}\hat{H}^{\prime}+
f^{\prime}_{4}\hat{\eta}^{\prime}\hat{\phi}^{\prime}\hat{H}^{\prime} 
\nonumber \\ &+&
f^{\prime}_{5}\hat{\phi}^{\prime}\hat{\phi}^{\prime}\hat{H}^{\prime}+
f^{\prime}_{6}\hat{\chi}^{\prime}\hat{\rho}^{\prime}\hat{H}^{\prime}+
\sum_{i=1}^{3}\kappa_{1i} \hat{Q}_{1L} \hat{\eta}^{\prime} \hat{u}^{c}_{iL}+
\kappa^{\prime}_{1} \hat{Q}_{1L} \hat{\eta}^{\prime} \hat{u}^{\prime c}_{L}+
\sum_{i=1}^{3}\kappa_{2i} \hat{Q}_{1L} \hat{\phi}^{\prime} \hat{u}^{c}_{iL}+
\kappa^{\prime}_{2}\hat{Q}_{1L} \hat{\phi}^{\prime} \hat{u}^{\prime c}_{L} \nonumber \\ &+&
\sum_{i=1}^{3}\kappa_{3i} \hat{Q}_{1L} \hat{\rho}^{\prime} \hat{d}^{c}_{iL}+
\sum^{2}_{\beta=1}\kappa_{3 \beta} \hat{Q}_{1L} \hat{\rho}^{\prime} \hat{d}^{\prime c}_{ \beta L}+
\kappa_{4} \hat{Q}_{1L} \hat{\chi}^{\prime} \hat{J}^{c}_{L}+
\sum_{\alpha=2}^{3}\sum_{i=1}^{3}\kappa_{5 \alpha i} \hat{Q}_{\alpha L} \hat{\rho} \hat{u}^{c}_{iL}+
\sum_{\alpha=2}^{3}\kappa^{\prime}_{5 \alpha} \hat{Q}_{\alpha L} \hat{\rho} \hat{u}^{\prime c}_{L} 
\nonumber \\ &+&
\sum_{\alpha=2}^{3}\sum_{i=1}^{3}\kappa_{6 \alpha i} \hat{Q}_{\alpha L} \hat{\eta} \hat{d}^{c}_{iL}+
\sum_{\alpha=2}^{3}\sum^{2}_{\beta=1}\kappa^{\prime}_{6 \alpha \beta} \hat{Q}_{\alpha L} \hat{\eta} \hat{d}^{\prime c}_{ \beta L}+
\sum_{\alpha=2}^{3}\sum_{i=1}^{3}\kappa_{7 \alpha i} \hat{Q}_{\alpha L} \hat{\phi} \hat{d}^{c}_{iL} \nonumber \\ &+&
\sum_{\alpha=2}^{3}\sum^{2}_{\beta=1}\kappa^{\prime}_{7 \alpha \beta} \hat{Q}_{\alpha L} \hat{\phi} \hat{d}^{\prime c}_{ \beta L}+ 
\sum_{\alpha=2}^{3}\sum_{\beta=1}^{2}\kappa_{8 \alpha \beta} \hat{Q}_{\alpha L} \hat{\chi} \hat{j}^{c}_{\beta L} \nonumber \\ &+& 
\sum_{\alpha=2}^{3}\sum_{a=1}^{3}\sum_{i=1}^{3}\kappa_{9 \alpha ai} \hat{Q}_{\alpha L} \hat{L}_{aL} \hat{d}^{c}_{iL}+ 
\sum_{a=1}^{3}\sum_{\alpha =1}^{2}\sum_{\beta =1}^{2}
\kappa^{\prime}_{9a \alpha \beta}\hat{L}_{aL} \hat{Q}_{\alpha L} \hat{d}^{\prime c}_{\beta L}+
\sum_{i=1}^{3}\sum_{j=1}^{3}\sum_{k=1}^{3}\xi_{1ijk} \hat{d}^{c}_{iL} 
\hat{d}^{c}_{jL} \hat{u}^{c}_{kL} \nonumber \\ &+&
\sum_{i=1}^{3}\sum_{j=1}^{3}\xi_{2ij} \hat{d}^{c}_{iL} 
\hat{d}^{c}_{jL} \hat{u}^{\prime c}_{L}+
\sum_{i=1}^{3}\sum_{\beta =1}^{2}\sum_{j=1}^{3}\xi_{3i \beta j} 
\hat{d}^{c}_{iL} \hat{d}^{\prime c}_{\beta L} \hat{u}^{c}_{jL}+
\sum_{i=1}^{3}\sum_{\beta =1}^{2}\xi_{4i \beta j} 
\hat{d}^{c}_{iL} \hat{d}^{\prime c}_{\beta L} \hat{u}^{\prime c}_{L} \nonumber \\ &+&
\sum_{\alpha =1}^{2}\sum_{\beta =1}^{2}\sum_{i=1}^{3}
\xi_{6 \alpha \beta i} \hat{d}^{\prime c}_{\alpha L}
\hat{d}^{\prime c}_{\beta L} \hat{u}^{c}_{iL}+
\sum_{\alpha =1}^{2}\sum_{\beta =1}^{2}\xi_{6 \alpha \beta} 
\hat{d}^{\prime c}_{\alpha L}\hat{d}^{\prime c}_{\beta L}
\hat{u}^{\prime c}_{L}+
\sum_{i=1}^{3}\sum_{j=1}^{3}\sum_{\beta=1}^{2}\xi_{7ij \beta} \hat{u}^{c}_{iL} 
\hat{u}^{c}_{jL} \hat{j}^{c}_{\beta L} \nonumber \\ &+&
\sum_{i=1}^{3}\sum_{\beta =1}^{2}\xi_{8i \beta} \hat{u}^{c}_{iL}\hat{u}^{\prime c}_{L}\hat{j}^{c}_{\beta L}+
\sum_{\beta =1}^{2}\xi_{9 \beta} \hat{u}^{\prime c}_{L}\hat{u}^{\prime c}_{L}\hat{j}^{c}_{\beta L}+
\sum_{i=1}^{3}\sum_{\beta=1}^{2}\xi_{10i \beta} \hat{d}^{c}_{iL} 
\hat{J}^{c}_{L} \hat{j}^{c}_{\beta L}. \nonumber \\
\label{sp3}
\end{eqnarray}

In the next subsection, we will show  that it is possible to define the $R$-parity symmetry, the phenomenology 
of this model with $R$-parity conserved has similar features to that of the 
$R$-conserving MSSM: the supersymmetric particles are pair-produced and the
lightest neutralino is the lightest supersymmetric particle (LSP), see in (subsection \ref{sec:wrc}). Then in 
(subsection \ref{sec:wrv}) we will show that there are terms that will induce mass to the neutrinos of the model.

\subsubsection{Discrete R-Parity in SUSY341}

To get the terms that are invariant under discrete $R$-parity in the superpotential, 
we need to check the following condition \cite{dong}
\begin{equation} 
\int d^{2}\theta\;
\prod_{a}\Phi_{a}(x,\theta ,\bar{\theta}) \,\ , \hspace{2cm}
\mbox{if  } \sum_{a} n_{a} = 0. 
\label{invrparitydiscreta} 
\end{equation}

Choosing the following R-charges 
\begin{eqnarray}
n_{\eta^{\prime}}&=&n_{\phi}=n_{\rho}=n_{H^{\prime}}=1, \,\ 
n_{\eta}=n_{\phi}=n_{\rho^{\prime}}=n_{H}=-1, \nonumber \\
n_{L}&=&n_{Q_{1}}=n_{Q_{\alpha}}=n_{d_{i}}=n_{u^{\prime}}= \frac{1}{2}, \, 
n_{J}=n_{j}=- \frac{1}{2}, \nonumber \\
n_{u_{i}}&=&n_{d^{\prime}}=- \frac{3}{2}, \,\ n_{\chi}=n_{\chi^{\prime}}=0,
\label{rdiscsusy331rn} 
\end{eqnarray}
with this $R$-charge assignment, we get that all the usual particle in the 
341 model has $R$-charge equal one while their superpartner has $R$-charge 
opposite, as happen in the MSSM.

Next we will present some discussion about the physical consequences on the superpotential, and point 
some likeness with the MSSM phenomenology.

\subsubsection{$R$-parity conservation}
\label{sec:wrc}

The terms in the superpotential that satisfy the $R$-parity, $W_{2RC}+W_{3RC}$, are given by the following terms
\begin{eqnarray}
W_{2RC}&=&\mu_{ \eta} \hat{ \eta} \hat{ \eta}^{\prime}+
\mu_{ \phi} \hat{ \phi} \hat{ \phi}^{\prime}+
\mu_{ \rho} \hat{ \rho} \hat{ \rho}^{\prime}+ 
\mu_{ \chi} \hat{ \chi} \hat{ \chi}^{\prime}+
\mu_{H} \hat{H} \hat{H}^{\prime}, \nonumber \\
W_{3RC}&=& \sum_{a=1}^{3} \sum_{b=1}^{3}
\lambda_{2ab} \epsilon \hat{L}_{aL} \hat{L}_{bL} \hat{ \eta}+
\sum_{a=1}^{3} \sum_{b=1}^{3} 
\lambda_{4ab} \hat{L}_{aL} \hat{L}_{bL} \hat{H}+ 
f_{1} \epsilon \hat{ \rho} \hat{ \chi} \hat{ \eta}+
f_{6} \hat{ \chi} \hat{ \rho} \hat{H}+ 
f^{\prime}_{1}\epsilon \hat{\rho}^{\prime}\hat{\chi}^{\prime}\hat{\eta}^{\prime} \nonumber \\ &+&
f^{\prime}_{6}\hat{\chi}^{\prime}\hat{\rho}^{\prime}\hat{H}^{\prime}+  
\sum_{i=1}^{3}\kappa_{1i} \hat{Q}_{1L} \hat{\eta}^{\prime} \hat{u}^{c}_{iL}+
\kappa^{\prime}_{2} \hat{Q}_{1L} \hat{\phi}^{\prime} \hat{u}^{\prime c}_{L}+
\sum_{i=1}^{3}\kappa_{3i} \hat{Q}_{1L} \hat{\rho}^{\prime} \hat{d}^{c}_{iL}+
\kappa_{4} \hat{Q}_{1L} \hat{\chi}^{\prime} \hat{J}^{c}_{L} \nonumber \\ &+&
\sum_{\alpha=2}^{3}\sum_{i=1}^{3}\kappa_{5 \alpha i} \hat{Q}_{\alpha L} \hat{\rho} \hat{u}^{c}_{iL}+
\sum_{\alpha=2}^{3}\sum_{i=1}^{3}\kappa_{6 \alpha i} \hat{Q}_{\alpha L} \hat{\eta} \hat{d}^{c}_{iL}+
\sum_{\alpha=2}^{3}\sum^{2}_{\beta=1}\kappa^{\prime}_{7 \alpha \beta} \hat{Q}_{\alpha L} \hat{\phi} \hat{d}^{\prime c}_{ \beta L} 
\nonumber \\ &+&
\sum_{\alpha=2}^{3}\sum_{\beta=1}^{2}\kappa_{8 \alpha \beta} \hat{Q}_{\alpha L} \hat{\chi} \hat{j}^{c}_{\beta L}.
\label{sprc}
\end{eqnarray}
The first term in $W_{3RC}$, will leave some leptons massless and some other mass generate, because 
$L_{a}L_{b} \equiv 4 \otimes 4=6 \oplus 10$ and how this term is antisymmetric in the generation indices ($a,b$) it 
implies that the Yukawa coupling $\lambda_{2ab}$ is antisymmetric  matrix. We have three antisymmetric factors
hence only the antisymmetric part of the coupling
constants $\lambda_{2ab}$ gives a non-vanishing contribution
and the mass matrix has eigenvalues
$0, -M, M$, so that one of the leptons does not gain mass
and the other two are degenerate, at least at tree level. Due this fact the second term will generate masses to the charged 
leptons \cite{su4b}.

With this superpotential, it is possible to give mass to all charged fermions in the model but 
neutrinos remain massless. Using Eq.(\ref{vev}) in Eq.(\ref{sprc}), we get the following mass 
matrices in the quark sector:
\begin{equation}
\Gamma^u=\frac{1}{\sqrt2}\,\left(\begin{array}{ccc}
\kappa_{11}v^{\prime} & \kappa_{12}v^{\prime} &
\kappa_{13}v^{\prime}\\ 
\kappa_{521}u & \kappa_{522}u & \kappa_{523}u \\
\kappa_{531}u & \kappa_{532}u & \kappa_{533} u\\
\end{array}\right),
\label{qumm}
\end{equation}
for the $u$-quarks, and 
\begin{equation}
\Gamma^d=\frac{1}{\sqrt2}\,\left(\begin{array}{ccc}
\kappa_{31}u^{\prime} & \kappa_{32} u^{\prime} &
\kappa_{33}u^{\prime} \\ 
\kappa_{621} v  & \kappa_{622} v & \kappa_{623} v \\
 \kappa_{631} v& \kappa_{632} v & \kappa_{633} v\\
\end{array}\right),
\label{qdmm}
\end{equation}
for the $d$-quarks, and for the exotic quarks, $J$ and $j_{1,2}$,we have
$M_J=\kappa_{4}w^{\prime}/\sqrt{2}$ and 
\begin{equation}
\Gamma^j=\frac{w}{\sqrt2}\left(\begin{array}{cc}
\kappa_{821} & \kappa_{822} \\
\kappa_{831} & \kappa_{832}
\end{array}\right),
\label{qjmm}
\end{equation}
respectively. The quark $u^{\prime}$ get the following mass 
$M_{u^{\prime}}=\kappa^{\prime}_{2}z^{\prime}/\sqrt{2}$ while the quark 
$d^{\prime}$ has
\begin{equation}
\Gamma^{d^{\prime}}=\frac{z}{\sqrt2}\left(\begin{array}{cc}
\kappa^{\prime}_{721} & \kappa^{\prime}_{722} \\
\kappa^{\prime}_{731} & \kappa^{\prime}_{732}
\end{array}\right),
\label{qdpmm}
\end{equation}

From Eqs.(\ref{qumm}) and (\ref{qdmm}) we see that all the VEVs from the quartet and 
anti-quartet have to be different from zero in order to give mass to all quarks. Notice also that
the $u$-like and $d$-like mass matrices have no common VEVs. On the other
hand, the charged lepton mass matrix is already given by
$M^l_{ij}=x\lambda_{3ij}/\sqrt2$, where $x$ is the VEV of
the $\langle H^0_3 \rangle$ component of the anti-decuplet $H$ in
Eq.(\ref{sextet}). However, all other VEV from the anti-decouplet and from the 
$x^{\prime}$ can both be zero since
the decouplet $H^\prime$ does not couple to leptons at all.

\subsubsection{$R$-parity violation}
\label{sec:wrv}

While the $R$-parity violating terms are given by $W_{2RV}+W_{3RV}$, where
\begin{eqnarray}
W_{2RV}&=&\sum_{a=1}^{3}\mu_{0a}\hat{L}_{aL} \hat{ \eta}^{\prime}+
\sum_{a=1}^{3}\mu_{1a}\hat{L}_{aL} \hat{ \phi}^{\prime}+
\mu_{2} \hat{ \eta} \hat{ \phi}^{\prime}+
\mu_{3} \hat{ \phi} \hat{ \eta}^{\prime}, \nonumber \\
W_{3RV}&=& \sum_{a=1}^{3} \sum_{b=1}^{3} \sum_{c=1}^{3}
\lambda_{1abc} \epsilon \hat{L}_{aL} \hat{L}_{bL} \hat{L}_{cL}+
\sum_{a=1}^{3} \sum_{b=1}^{3}
\lambda_{3ab} \epsilon \hat{L}_{aL} \hat{L}_{bL} \hat{ \phi}+
+ \sum_{a=1}^{3} \lambda_{5a} \epsilon \hat{L}_{aL} \hat{\chi} \hat{\rho}+
f_{2} \epsilon \hat{ \rho} \hat{ \chi} \hat{ \phi} \nonumber \\ &+&
f_{3} \hat{ \eta} \hat{ \eta} \hat{H}+
f_{4} \hat{ \eta} \hat{ \phi} \hat{H}+
f_{5} \hat{ \phi} \hat{ \phi} \hat{H}+
f^{\prime}_{2}\epsilon \hat{\rho}^{\prime}\hat{\chi}^{\prime}\hat{\phi}^{\prime}+
f^{\prime}_{3}\hat{\eta}^{\prime}\hat{\eta}^{\prime}\hat{H}^{\prime}+
f^{\prime}_{4}\hat{\eta}^{\prime}\hat{\phi}^{\prime}\hat{H}^{\prime}+
f^{\prime}_{5}\hat{\phi}^{\prime}\hat{\phi}^{\prime}\hat{H}^{\prime} \nonumber \\ &+& 
\kappa^{\prime}_{1} \hat{Q}_{1L} \hat{\eta}^{\prime} \hat{u}^{\prime c}_{L}+
\sum_{i=1}^{3}\kappa_{2i} \hat{Q}_{1L} \hat{\phi}^{\prime} \hat{u}^{c}_{iL}+
\sum^{2}_{\beta=1}\kappa_{3 \beta} \hat{Q}_{1L} \hat{\rho}^{\prime} \hat{d}^{\prime c}_{ \beta L}+
\sum_{\alpha=2}^{3}\kappa^{\prime}_{5 \alpha} \hat{Q}_{\alpha L} \hat{\rho} \hat{u}^{\prime c}_{L} \nonumber \\ &+&
\sum_{\alpha=2}^{3}\sum^{2}_{\beta=1}\kappa^{\prime}_{6 \alpha \beta} \hat{Q}_{\alpha L} \hat{\eta} \hat{d}^{\prime c}_{ \beta L}+
\sum_{\alpha=2}^{3}\sum^{2}_{\beta=1}\kappa^{\prime}_{7 \alpha \beta} \hat{Q}_{\alpha L} \hat{\phi} \hat{d}^{\prime c}_{ \beta L}+ 
\sum_{\alpha=2}^{3}\sum_{a=1}^{3}\sum_{i=1}^{3}\kappa_{9 \alpha ai} \hat{Q}_{\alpha L} \hat{L}_{aL} \hat{d}^{c}_{iL} 
\nonumber \\ &+& 
\sum_{a=1}^{3}\sum_{\alpha =1}^{2}\sum_{\beta =1}^{2}
\kappa^{\prime}_{9a \alpha \beta}\hat{L}_{aL} \hat{Q}_{\alpha L} \hat{d}^{\prime c}_{\beta L}+
\sum_{i=1}^{3}\sum_{j=1}^{3}\sum_{k=1}^{3}\xi_{1ijk} \hat{d}^{c}_{iL} 
\hat{d}^{c}_{jL} \hat{u}^{c}_{kL}+
\sum_{i=1}^{3}\sum_{j=1}^{3}\xi_{2ij} \hat{d}^{c}_{iL} 
\hat{d}^{c}_{jL} \hat{u}^{\prime c}_{L} \nonumber \\ &+&
\sum_{i=1}^{3}\sum_{\beta =1}^{2}\sum_{j=1}^{3}\xi_{3i \beta j} 
\hat{d}^{c}_{iL} \hat{d}^{\prime c}_{\beta L} \hat{u}^{c}_{jL}+
\sum_{i=1}^{3}\sum_{\beta =1}^{2}\xi_{4i \beta j} 
\hat{d}^{c}_{iL} \hat{d}^{\prime c}_{\beta L} \hat{u}^{\prime c}_{L}+
\sum_{\alpha =1}^{2}\sum_{\beta =1}^{2}\sum_{i=1}^{3}
\xi_{5 \alpha \beta i} \hat{d}^{\prime c}_{\alpha L}
\hat{d}^{\prime c}_{\beta L} \hat{u}^{c}_{iL} \nonumber \\ &+&
\sum_{\alpha =1}^{2}\sum_{\beta =1}^{2}\xi_{6 \alpha \beta} 
\hat{d}^{\prime c}_{\alpha L}\hat{d}^{\prime c}_{\beta L}
\hat{u}^{\prime c}_{L}+
\sum_{i=1}^{3}\sum_{j=1}^{3}\sum_{\beta=1}^{2}\xi_{7ij \beta} \hat{u}^{c}_{iL} 
\hat{u}^{c}_{jL} \hat{j}^{c}_{\beta L}+
\sum_{i=1}^{3}\sum_{\beta =1}^{2}\xi_{8i \beta} \hat{u}^{c}_{iL}\hat{u}^{\prime c}_{L}\hat{j}^{c}_{\beta L} 
\nonumber \\ &+&
\sum_{\beta =1}^{2}\xi_{9 \beta} \hat{u}^{\prime c}_{L}\hat{u}^{\prime c}_{L}\hat{j}^{c}_{\beta L}+
\sum_{i=1}^{3}\sum_{\beta=1}^{2}\xi_{10i \beta} \hat{d}^{c}_{iL} 
\hat{J}^{c}_{L} \hat{j}^{c}_{\beta L}. 
\label{sprv}
\end{eqnarray}

The superpotential give in Eq.(\ref{sprv}) provide us the mass terms for leptons and 
higgsinos
\begin{equation}
- \frac{\mu_{0a}}{2}L_{aL} \tilde{ \eta}^{\prime}-
\frac{\mu_{1a}}{2}L_{aL} \tilde{ \phi}^{\prime}- 
\frac{\mu_{2}}{2} \tilde{ \eta} \tilde{ \phi}^{\prime}-
\frac{\mu_{3}}{2} \tilde{ \phi} \tilde{ \eta}^{\prime}-
\frac{\lambda_{5a}}{3}(L_{aL} \tilde{\chi}\rho+ \tilde{\rho}l_{aL}\chi)+hc,
\end{equation}
wchich will induce mass for thre neutrinos as was show in \cite{lepmass,dong}.

However the $R$-violating interactions can give the correct masses to 
$e,\mu$ and $\tau$, even without a decouplet in the same way as happened 
in thew minimal supersymmetric 331 model, \cite{lepmass}, where only the mixing 
between the higgsinos with the leptons reproduce the mass spectrum of the physical 
charged leptons.\footnote{Work in progress.}. 

By another hand, the superpotential give at Eq.(\ref{sprv}) will also induce the proton 
decay. The terms proportional to $\kappa_{9},\xi_{1},\xi_{3}$ would produce the dominant 
decay $p \rightarrow \pi^{0}e^{+}$, shown in Figs.(\ref{f3},\ref{f8}).

\begin{figure}[t]
\begin{center}
\begin{picture}(150,100)(0,0)
\SetWidth{1.2}
\ArrowLine(10,10)(50,50)
\Text(10,10)[r]{$u_{1}$}
\ArrowLine(10,90)(50,50)
\Text(10,90)[r]{$d_{1}$}
\Vertex(50,50){2.0}
\Text(30,50)[]{$\xi_{1}$}
\DashLine(100,50)(50,50){3}
\Vertex(100,50){2.0}
\Text(115,50)[]{$\kappa_{9}$}
\Text(75,58)[b]{$\tilde{d}^{c}_{2},\tilde{d}^{c}_{3}$}
\ArrowLine(140,90)(100,50)
\Text(145,90)[l]{$u^{c}_{1},u^{c}_{2}$}
\ArrowLine(140,10)(100,50)
\Text(145,10)[l]{$l^{+}_{1},l^{+}_{2},l^{+}_{3}$}
\end{picture}\\ 
{\sl Proton Decay in charged leptons in the SUSY341 exchanging $\tilde{d}$.}
\end{center}
\label{f3} 
\end{figure}

\begin{figure}[t]
\begin{center}
\begin{picture}(220,100)(0,0)
\SetWidth{1.1}
\ArrowLine(10,10)(50,50)
\Text(10,10)[r]{$u_{1}$}
\ArrowLine(10,90)(50,50)
\Text(10,90)[r]{$d_{1}$}
\Vertex(50,50){2.0}
\Text(30,50)[]{$\xi_{1}$}
\DashLine(90,50)(50,50){3}
\Vertex(90,50){2.0}
\Text(80,40)[]{$g$}
\Text(70,55)[b]{$\tilde{d}_{2}^{c}$}
\ArrowLine(90,90)(90,50)
\Text(95,90)[l]{$d^{c}_{2}$}
\Line(90,50)(130,50)
\Text(110,45)[t]{$\tilde{\gamma}, \tilde{Z}, \tilde{Z}^{\prime}$}
\Vertex(130,50){2.0}
\Text(140,60)[]{$g$}
\ArrowLine(130,50)(130,90)
\Text(160,90)[b]{$d_{1},d_{2}$}
\DashLine(170,50)(130,50){3}
\Vertex(170,50){2.0}
\Text(190,50)[]{$\xi_{3}$}
\Text(160,30)[b]{$\tilde{d}^{\prime c}_{1},\tilde{d}^{\prime c}_{2}$}
\ArrowLine(210,90)(170,50)
\Text(220,90)[l]{$u^{c}_{1},u^{c}_{2}$}
\ArrowLine(210,10)(170,50)
\Text(220,10)[l]{$d^{c}_{1},d^{c}_{2}$}
\end{picture}\\ 
{\sl Proton Decay in charged leptons in the SUSY341 exchanging $\tilde{d}^{\prime}$.}
\end{center}
\label{f8}
\end{figure}

Remember that in the MSSM $\tilde{m}_{d} \propto m_{d}$, and due the fact that 
$m_{d^{\prime}}>m_{d}$ then it implies that $\tilde{m}_{{d}^{\prime}}>\tilde{m}_{d}$. Due this 
fact we can negligible the contribution coming from the Fig.(\ref{f8}) and imposing
\begin{equation}
\kappa_{4 \alpha 11}\xi_{1 \alpha 11}< 5,29 \times 10^{-25},
\end{equation}
we will get that $\tau(p \to e \pi)
>1.6 \times 10^{33}$ years \cite{pallong} \footnote{The decay mode 
$p \to K^{+}e^{\mp}\mu^{\pm}\nu_{\tau}$ was considered in Ref~\cite{pallong} is also present in this model.}.

\subsection{The soft terms}
\label{subsec:softterm}

 The soft terms  can be written as  
\begin{equation}
{\cal L}_{\mbox{soft}}={\cal L}_{GMT}+
{\cal L}^{\mbox{soft}}_{\mbox{Scalar}}+{\cal L}_{SMT},
\label{soft}
\end{equation}
where
\begin{eqnarray}
{\cal L}_{GMT}&=&- \frac{1}{2} \left[ m_{ \lambda_{C}} \sum_{b=1}^{8} 
\left( \lambda^{b}_{C} \lambda^{b}_{C} \right) +m_{ \lambda} \sum_{b=1}^{15} 
\left( \lambda^{b}_{A} \lambda^{b}_{A} \right) +  m^{ \prime} \lambda_{B} \lambda_{B}+H.c. \right] , \quad
\label{gmt}
\end{eqnarray}
give mass to the boson superpartners and
\begin{eqnarray}
{\cal L}^{\mbox{soft}}_{\mbox{Scalar}}&=&
-m^2_{ \eta}\eta^{ \dagger}\eta-m^{2}_{\phi}\phi^{ \dagger}\phi  -
m^2_{ \rho}\rho^{ \dagger}\rho-m^2_{ \chi}\chi^{ \dagger}\chi - 
m^{2}_{H}H^{ \dagger}H-m^{2}_{1}\phi^{ \dagger}\eta^{\prime} -
m^2_{\eta^{\prime}}\eta^{\prime \dagger}\eta^{\prime}-
m^{2}_{\phi^{\prime}}\phi^{\prime \dagger}\phi^{\prime} \nonumber \\ 
&-&
m^2_{\rho^{\prime}}\rho^{\prime \dagger}\rho^{\prime}-
m^2_{\chi^{\prime}}\chi^{\prime \dagger}\chi^{\prime}-
m^{2}_{H^{\prime}}H^{\prime \dagger}H^{\prime}-
m^{2}_{2}\phi^{\prime \dagger}\eta +[
k_{1} \epsilon_{ijk}\rho_{i}\chi_{j}\eta_{k}+ 
k_{2}\chi \rho H +
k^{\prime}_{1} \epsilon_{ijk}\rho^{\prime}_{i}\chi^{\prime}_{j}\eta^{\prime}_{k} 
\nonumber \\ &+& 
k^{\prime}_{2}\chi^{\prime} \rho^{\prime} H^{\prime}].
\label{potencial}
\end{eqnarray}
in order to give mass to the scalars, we have omitting the sum upon repeated
indices, $i,j,k=1,2,3$ , and we only write the terms that respect $R$-parity 
see Eq.(\ref{sprc}). Finally, we have to add
\begin{eqnarray}
-{\cal L}_{SMT}&=& m^{2}_{L_a} \tilde{L}^{\dagger}_{aL}
\tilde{L}_{aL}+
 m_{la}^{2}\tilde{l}^{c \dagger}_{aL} \tilde{l}^{c}_{aL}+ 
m^2_{Q_3} 
\tilde{Q}^{\dagger}_{1L} \tilde{Q}_{1L}+
m_{Q_{\alpha L}}^{2} 
\tilde{Q}^{\dagger}_{\alpha L} \tilde{Q}_{\alpha L}+ m_{u_{i}}^2 
\tilde{u}^{c \dagger}_{iL} \tilde{u}^{c}_{iL}+ m_{d_{i}}^2 
\tilde{d}^{c \dagger}_{iL} \tilde{d}^{c}_{iL} \nonumber \\
&+&m_{u'}^{2} \tilde{u}^{\prime c \dagger}_L \tilde{u}^{\prime c}_L 
+ m_{d_{ \alpha}^{'}}^{2} 
\tilde{d}^{\prime c \dagger}_{ \beta L} \tilde{d}^{\prime c}_{ \beta L} +
m_{J}^{2} \tilde{J}^{\dagger} \tilde{J}-
\sum_\beta \tilde{j}^{\dagger}_{\beta}
m_{j_{ \beta}}^{2} \tilde{j}_{ \beta}+[
\upsilon_{1ab} \epsilon \tilde{L}_{aL} \tilde{L}_{bL} \eta+
\upsilon_{2ab} \tilde{L}_{aL} \tilde{L}_{bL} H \nonumber \\ &+& 
\vartheta_{1\alpha i} \tilde{Q}_{1L} \eta^{\prime} \tilde{u}^{c}_{iL}+
\vartheta_{2\alpha} \tilde{Q}_{1L} \phi^{\prime} \tilde{u}^{\prime c}_{L}+
\vartheta_{3i} \tilde{Q}_{1L} \rho^{\prime} \tilde{d}^{c}_{iL}+
\vartheta_{4} \tilde{Q}_{1L} \chi^{\prime} \tilde{J}^{c}_{L}+
\vartheta_{5 \alpha i} \tilde{Q}_{\alpha L} \rho \tilde{u}^{c}_{iL} \nonumber \\ &+&
\vartheta_{6\alpha i} \tilde{Q}_{\alpha L} \eta \tilde{d}^{c}_{iL}+ 
\vartheta_{7\alpha}\tilde{Q}_{\alpha L}\phi\tilde{d}^{\prime c}_{L}+
\vartheta_{8\alpha \beta} \tilde{Q}_{\alpha L} \chi \tilde{j}^{c}_{\beta L}]
\label{soft}
\end{eqnarray}

The pattern of the symmetry breaking in this  model is given by

\begin{eqnarray}
&\mbox{susy341}& \stackrel{{\cal L}_{soft}}{\longmapsto}
\mbox{SU(3)}_C\ \otimes \ \mbox{SU(4)}_L\otimes \mbox{U(1)}_N
\stackrel{\langle\chi\rangle \langle
\chi^{\prime}\rangle}{\longmapsto} \mbox{SU(3)}_C \ \otimes \
\mbox{SU(2)}_L\otimes
\mbox{U(1)}_Y \nonumber \\
&\stackrel{\langle\rho,\eta,\phi,H \rho^{\prime}\eta^{\prime},\phi^{\prime},H^{\prime}\rangle}{\longmapsto}&
\mbox{SU(3)}_C \ \otimes \ \mbox{U(1)}_Q.
\end{eqnarray}

\section{Double Chargino and Neutralino Production $e^{-}e^{-} \to \tilde{ \chi}^{--} \tilde{ \chi}^{0}$}
\label{sec:double}

Models with  $SU(3)$ (or $SU(4)$) electroweak symmetry and left-right symmetry (LR) may have doubly charged 
vector bosons and scalars, respectivelly. This means that in some supersymmetric extensions of these kind 
of models we will have double charged charginos \cite{mcr}. On the table, Tab.(\ref{t2}), 
\begin{table}
\begin{center}
\begin{tabular} {|c|c|}\hline
model    & charginos and neutralinos \\ \hline 
MSSM \cite{mssm} & $\tilde{\chi}^{\pm}(2) \,\ \tilde{\chi}^{0}(4)$ \\ \hline
NMSSM \cite{dress}& $\tilde{\chi}^{\pm}(2) \,\ \tilde{\chi}^{0}(5)$ \\ \hline
SUSYLRT \cite{susylr}& $\tilde{\chi}^{\pm \pm}(1) \,\ \tilde{\chi}^{\pm}(5) \,\ 
\tilde{\chi}^{0}(9)$ \\ \hline
SUSYLRD \cite{doublet}& $\tilde{\chi}^{\pm}(6) \,\ \tilde{\chi}^{0}(11)$ \\ \hline
MSUSY331 \cite{331susy} &   $\tilde{\chi}^{\pm \pm}(5) \,\ \tilde{\chi}^{\pm}(8) \,\  
\tilde{ \chi}^{0}(13)$ \\ \hline
SUSY331RN \cite{331susy2} & $\tilde{ \chi}^{\pm}(6) \,\ \tilde{ \chi}^{0}(15)$  \\ \hline
SUSY341 & $\tilde{\chi}^{\pm \pm}(5) \,\ \tilde{\chi}^{\pm}(16) \,\  \tilde{\chi}^{0}(25)$ \\ \hline 
\end{tabular}
\end{center}
\caption{Spectrum of Charginos and Neutralinos in several SUSY models}
\label{t2}
\end{table}
we show the possible states in the supersymmetric models, in parenthesis we show the number of 
states that they appear. Therefore we can distinguish the differents models with base in the numbers 
of particles. In Left-Right Supersymmetric Model (SUSYLRT) the doubly charged higgsinos do not mix with gauginos.
The production of a double charged higgsino was studied in \cite{LR}.

Because of low level of SM backgrounds, the total cross section 
$\sigma \approx 10^{-3}nb$ at $\sqrt{s}=500GeV$ \cite{assi3}, $e^-e^-$ 
collisions are a good reaction for discovering and investigating new physics 
at linear colliders. With this process is possible to study reactions that 
violate both lepton and/or fermion number, and this kind of reaction are expected in 
supersymmetric models, as we will briefelly present next.

In the minimal supersymmetric 331 model the chargino (neutralino) base is given by
\begin{eqnarray}
\psi^{++}&=& \left( \begin{array}{rrrrr}
-i \lambda^{++}_{U}&
\tilde{\rho}^{++}&
\tilde{\chi}^{ \prime ++}&
\tilde{H}_1^{++}&
\tilde{H}_2^{ \prime ++}
\end{array}
\right)^t \nonumber \\
\psi^{+}&=& \left( \begin{array}{rrrrrrrr}
-i \lambda^{+}_{W}&
-i \lambda^{+}_{V}&
\tilde{\eta}_{1}^{ \prime +}&
\tilde{\eta}_{2}^{+}&
\tilde{\rho}^{+}&
\tilde{\chi}^{ \prime +}&
\tilde{h}_{1}^{ \prime +}&
\tilde{h}_{2}^{+}
\end{array}
\right)^t \nonumber \\ 
\Psi^{0}&=& \left( \begin{array}{rrrrrrrrrrrrr}
-i \lambda^{3}_{A} &
-i \lambda^{8}_{A}&
-i \lambda_{B}&
\tilde{\eta}^{0}&
\tilde{\eta}^{ \prime 0}&
\tilde{\rho}^{0}&
\tilde{\rho}^{ \prime 0}&
\tilde{\chi}^{0}&
\tilde{\chi}^{ \prime 0}&
\tilde{\sigma}_{1}^{0}&
\tilde{\sigma}_{1}^{ \prime 0}&
\tilde{\sigma}_{2}^{0}&
\tilde{\sigma}_{2}^{ \prime 0}
\end{array}
\right)^t \nonumber \\
\end{eqnarray}
while the base on the susy341 model is
\begin{eqnarray}
\psi^{++}&=& \left( \begin{array}{rrrrr}
-i \lambda^{++}_{U} &
\tilde{\rho}^{++} &
\tilde{\chi}^{\prime ++} &
\tilde{H}_1^{++} &
\tilde{H}_2^{\prime ++}
\end{array}
\right) \nonumber \\
\psi^{+}&=& \left( \begin{array}{r}
-i \lambda^{+}_{W} \\
-i \lambda_{V_{1}}^{+}\\
-i \lambda_{V_{2}}^{+}\\
-i \lambda_{V_{3}}^{+}\\
\eta^{\prime +}_{1} \\
\eta^{+}_{2} \\
\phi^{\prime +}_{1} \\
\phi^{+}_{2} \\
\rho^{+}_{1} \\
\rho^{+}_{2} \\
\chi^{\prime +}_{1} \\
\chi^{\prime +}_{2} \\
H^{+}_{1} \\
H^{\prime +}_{2} \\
H^{+}_{3} \\
H^{\prime +}_{4}
\end{array}
\right) \quad
\Psi^{0}= \left( \begin{array}{r}
-i \lambda^{3}_{A} \\
-i \lambda_{X^{0}} \\
-i \lambda_{X^{0*}} \\
-i \lambda^{8}_{A} \\
-i \lambda^{15}_{A} \\
-i \lambda_{B} \\
\tilde{\eta}^{0}_{1} \\
\tilde{\eta}^{0}_{2} \\
\tilde{\eta}^{\prime 0}_{1} \\
\tilde{\eta}^{\prime 0}_{2} \\
\tilde{\phi}^{0}_{1} \\
\tilde{\phi}^{0}_{2} \\
\tilde{\phi}^{\prime 0}_{1} \\
\tilde{\phi}^{\prime 0}_{2} \\
\tilde{\rho}^{0} \\
\tilde{\rho}^{\prime 0} \\
\tilde{\chi}^{0} \\
\tilde{\chi}^{\prime 0} \\
\tilde{H}_{1}^{0} \\
\tilde{H}_{2}^{0} \\
\tilde{H}_{3}^{0} \\
\tilde{H}_{4}^{0} \\
\tilde{H}_{1}^{\prime 0} \\
\tilde{H}_{2}^{\prime 0} \\
\tilde{H}_{3}^{\prime 0} \\
\tilde{H}_{4}^{\prime 0}
\end{array}
\right) 
\end{eqnarray}

The interaction lagrangian in the supersymmetric 331 model is presented at Appendix A of \cite{mcr}, the respective 
lagrangian interaction in the susy341 model is given by
\begin{eqnarray}
{\cal L}^{\mbox{lep}}_{llV}&=&\frac{g}{2} \bar{L}\bar\sigma^{m}\lambda^{a} LV^{a}_{m}, \nonumber \\
{\cal L}^{\mbox{lep}}_{l \tilde{l} \tilde{V}}&=&- \frac{ig}{ \sqrt{2}}
( \bar{L}\lambda^{a}\tilde{L}\bar{\lambda}^{a}_{A}- 
\bar{\tilde{L}}\lambda^{a}L\lambda^{a}_{A}). \nonumber \\
{\cal L}^{\mbox{gauge}}_{ \lambda \lambda V}&=&-igf^{abc}\bar{\lambda}^{a}_{A}
\lambda^{b}_{A} \sigma^{m}V^{c}_{m}, \nonumber \\
{\cal L}^{\mbox{Scalar}}_{ \tilde{H} \tilde{H}V}&=& \frac{g}{2} \left[ 
\bar{\tilde{\eta}}\bar\sigma^{m}\lambda^{a} \tilde{\eta}+ 
\bar{\tilde{\phi}}\bar\sigma^{m}\lambda^{a} \tilde{\phi}+
\bar{\tilde{\rho}}\bar\sigma^{m}\lambda^{a} \tilde{\rho}+
\bar{\tilde{\chi}}\bar\sigma^{m}\lambda^{a} \tilde{\chi}+ 
\bar{\tilde{H}}\bar\sigma^{m}\lambda^{a} \tilde{H} \right. \nonumber \\
&-& \left. 
\bar{\tilde{\eta}}^{\prime}\bar\sigma^{m}\lambda^{* a} \tilde{\eta}^{\prime}-
\bar{\tilde{\phi}}^{\prime}\bar\sigma^{m}\lambda^{* a} \tilde{\phi}^{\prime}- 
\bar{\tilde{\rho}}^{\prime}\bar\sigma^{m}\lambda^{* a} \tilde{\rho}^{\prime}-
\bar{\tilde{\chi}}^{\prime}\bar\sigma^{m}\lambda^{* a} \tilde{\chi}^{\prime}- 
\bar{\tilde{H}}^{\prime}\bar\sigma^{m}\lambda^{* a} \tilde{H}^{\prime}  
\right]V^{a}_{m} \nonumber \\
&+& 
\frac{g^{ \prime}}{2} \left[ \bar{\tilde{\rho}}\bar{\sigma}^{m}\tilde{\rho}-
\bar{\tilde{\chi}}\bar{\sigma}^{m}\tilde{\chi}-  
\bar{\tilde{\rho}}^{\prime}\bar{\sigma}^{m}\tilde{\rho}^{\prime}+
\bar{\tilde{\chi}}^{\prime}\bar{\sigma}^{m}\tilde{\chi}^{\prime} 
\right]V_{m}.
\end{eqnarray}
Comparing both lagrangian, we notice that both have the same structure. Terefore, the 
Feynman rules in both model are the same, and they are given in Table \ref{t1} we have defined the following operators:
\begin{eqnarray}
O^{1}_{ij}&=&A^{*}_{i1}(\sqrt{3}N_{j2}-N_{j1})+ \sqrt{2}A^{*}_{i2}N_{j5}+
A^{*}_{i5}N_{j8}, \nonumber \\
O^{2}_{ij}&=&- \left( D^{*}_{i1}E_{j2}-D^{*}_{i2}E_{j1}+D^{*}_{i4}E_{j3}+
\frac{1}{2}D^{*}_{i7}E_{j8} \right).
\end{eqnarray} 
This implies new interactions that are not present in the 
MSSM, for instance: $\tilde{\chi}^{--}\tilde{\chi}^0U^{++}$,
$\tilde{\chi}^-\tilde{\chi}^-U^{++}$, 
$\tilde{l}^-l^-\tilde{\chi}^{++}$ where $\tilde{\chi}^{++}$ denotes
any doubly charged chargino. 
Moreover, in the chargino production, besides the usual mechanism, we have 
additional contributions coming from the $U$-bilepton in the s-channel. 
Due to this fact we expect that there will be an enhancement in the cross 
section of production of these particles in $e^-e^-$ collisors, such as the
ILC.

\begin{table}
\begin{center}
\begin{tabular}{|c|c|} 
\hline
Vertices & Feynman rules   \\ \hline 
$l^-l^-U^{--}$ & $- \frac{ig}{\sqrt{2}} C \gamma^{m}L$ \\ \hline 
$\tilde{ \chi}^{--}_{j} \tilde{ \chi}^{0}_{i}U^{--}$ & 
$\frac{ig}{2} O^{1}_{ij}C \gamma^{m}R$ \\ \hline
$\tilde{ \chi}^{-}_{i} \tilde{ \chi}^{-}_{j}U^{--}$ &
$\frac{ig}{2}O^{2}_{ij}C \gamma^{m}R$ \\ \hline
$\tilde{l}^{-}_{1}l^{-} \tilde{\chi}^{--}_{i}$ & 
$- 2i \lambda_3A_{i5} \sin \theta_{f}R$ \\ \hline
$\tilde{l}^{-}_{2}l^{-} \tilde{\chi}^{--}_{i}$ &
$- 2i \lambda_3A_{i5} \cos \theta_{f}R$ \\ \hline
$\tilde{l}^{-}_{1}l^{-} \tilde{\chi}^{0}_{i}$ &
$i \left[ g \left( 
\frac{N_{i1}}{\sqrt{2}}+ \frac{N_{i2}}{\sqrt{6}} \right) \cos \theta_{f}R- 
\lambda_{3} \frac{2}{\sqrt{2}} \sin \theta_{f}N_{i8}R \right]$ \\ \hline
$\tilde{l}^{-}_{2}l^{-} \tilde{\chi}^{0}_{i}$ &
$i \left[ g \left( 
\frac{N_{i1}}{\sqrt{2}}+ \frac{N_{i2}}{\sqrt{6}} \right) \sin \theta_{f}R+ 
\lambda_{3} \frac{2}{\sqrt{2}} \cos \theta_{f}N_{i8}R \right]$ \\ \hline
$\tilde{l}^{+}_{1}l^{-} \tilde{\chi}^{0}_{i}$ &
$i \left[ g \left( 
\frac{N^*_{i1}}{\sqrt{2}}+ \frac{N^*_{i2}}{\sqrt{6}} \right) \cos \theta_{f}L- 
\lambda_{3} \frac{2}{\sqrt{2}} \sin \theta_{f}N^*_{i8}L \right]$ \\ \hline
$\tilde{l}^{+}_{2}l^{-} \tilde{\chi}^{0}_{i}$ &
$i \left[ g \left( 
\frac{N^*_{i1}}{\sqrt{2}}+ \frac{N^*_{i2}}{\sqrt{6}} \right) \sin \theta_{f}L+ 
\lambda_{3} \frac{2}{\sqrt{2}} \cos \theta_{f}N^*_{i8}L \right]$ \\ \hline
$\tilde{l}^{+}_{1}l^{-} \tilde{\chi}^{--}_{i}$ &
$-igA_{i1} \sin \theta_{f}RC$ \\ \hline
$\tilde{l}^{+}_{2}l^{-} \tilde{\chi}^{--}_{i}$ &
$-igA_{i1} \cos \theta_{f}RC$ \\ \hline
$\tilde{\nu_{l}}l^{-} \tilde{\chi}^{-}_{i}$ &
$-i \lambda_{3} \frac{2}{\sqrt{2}}D^{*}_{i7}L$ \\ \hline
$\tilde{\nu_{l}}^{*}l^{-} \tilde{\chi}^{-}_{i}$ &
$-i \lambda_{3} \frac{2}{\sqrt{2}}D_{i7}R$ \\ \hline
\end{tabular}
\end{center}
\caption{Feynman rules derived from susy341 model in the same way as done in \cite{mcr}.}
\label{t1}
\end{table}

Therefore the result to the light 
double chargino production are the same as presented in \cite{mcr}. We must remember that the susy341 model 
differ from the minimal supersymmetric 331 model in the number of neutralinos. On 
this way we can distuinguish between both models.

In the future, we want to compare the results about the double chargino production 
on SUSYLRT, SUSY331 and SUSY341, bacause this kind of phenomenology was not so much studied 
in the literature and it can be very nice signal to new physics.. We believe that these new states can be 
discovered, if they really exist, in linear colliders ILC.

\section{Conclusions}
\label{sec:con}

We have built the complete supersymmetric version of the 3-4-1 model of
Ref.~\cite{su4b}. 
  
From the phenomenological point of view there are several possibilities.
Since it is possible to define the $R$-parity symmetry, the phenomenology 
of this model with $R$-parity conserved has similar features to that of the 
$R$-conserving MSSM: the supersymmetric particles are pair-produced and the
lightest neutralino is the lightest supersymmetric particle (LSP).

While in the case that $R$-parity is not conserved we can induce masses to neutrinos 
of the model in the same way as in the MSSM. We also studied the proton 
decay problem in this model, and we show that it is in asgreement with the experimental data.

\begin{center}
{\Large {\bf Acknowledgments}}
\end{center} 
This work was supported by Conselho Nacional de Ci\^encia e Tecnologia (CNPq) 
under the processes 309564/2006-9.

\end{document}